\documentclass[aps,prd,preprint,nofootinbib]{revtex4}
\usepackage{amsfonts}
\usepackage{mathrsfs}
\usepackage{graphicx}
\usepackage{amsmath}
\usepackage{amssymb}
\usepackage{epsfig}
\usepackage{graphicx}
\newcommand{\bqa}{\begin{eqnarray}}
\newcommand{\eqa}{\end{eqnarray}}
\newcommand{\beq}{\begin{equation}}
\newcommand{\eeq}{\end{equation}}
\graphicspath{{fig/}{dia/}} \DeclareGraphicsExtensions{.eps}
\begin{document}
\baselineskip 20pt
%%%%%%%%%%%%%%%%%%%%%%%%%%%%%%%%%%%%%%%%%%%%%%%%%%%%%%%%%%%%%%%%%%%%%%
\title{NLO QCD Corrections for $J/\psi+ c + \bar{c}$ Production in Photon-Photon Collision}

\author{\vspace{1cm} Zi-Qiang Chen$^1$\footnote{
chenziqiang13@mails.ucas.ac.cn}, Long-Bin Chen$^{1,2}$\footnote{
chenlongbin@ucas.ac.cn}  and Cong-Feng
Qiao$^{1,2}$\footnote{qiaocf@ucas.ac.cn, corresponding author} \\}

\affiliation{$^1$ School of Physics, University of Chinese Academy of
Sciences,  Yuquan Road 19A, Beijing 100049, China}

\affiliation{$^2$ CAS Center for Excellence in Particle Physics, Beijing 100049, China\vspace{0.6cm}}

\begin{abstract}

The $\gamma+\gamma\rightarrow J/\psi+c+\bar{c}$ inclusive process is an extremely important subprocess in $J/\psi$ production via photon-photon scattering, like at LEP\uppercase\expandafter{\romannumeral2} or various types future electron-positron colliders. In this work we perform the next-to-leading(NLO) QCD corrections to this process in the framework of non-relativistic QCD(NRQCD) factorization formalism, the first NLO calculation for two projectiles to 3-body quarkonium inclusive production process. By setting the center-of-mass energy at LEP\uppercase\expandafter{\romannumeral2}, the $\sqrt{s}=197$ GeV, we conduct analyses of the $p_t^2$ distribution and total cross section of this process at the NLO accuracy. It turns out that the total cross section is moderately enhanced by the NLO correction with a $K$ factor of about 1.46, and hence the discrepancy between DELPHI data and color-singlet(CS) calculation is reduced while the color-octet(CO) contributions are still inevitable at this order. At the future Circular Electron-Positron Collider(CEPC), the NLO corrections are found to be  more significant, with a $K$ factor of about 1.76.

\vspace {5mm} \noindent {PACS number(s): 13.66.Bc, 12.38.Bx,
14.40.Pq}
%\noindent {\bf Keywords:}
\end{abstract}
\maketitle
%%%%%%%%%%%%%%%%%%%%%%%%%%%%%%%%%%%%%%%%%%%%%%%%%%%%%%%%%%%%%%%%%%%%%%%%%%%%

The heavy quarkonium production and decay is an important topic to study in high energy physics, which may enrich our knowledge on quarkonium structure and the nonperturbative properties of QCD. The appearance of non-relativisitic QCD (NRQCD) factorization formalism makes the corresponding theoretical study on more solid footing \cite{NRQCD}.
Nevertheless, there still couple of open questions in the application of NRQCD, especially on quarkonium production mechanism. Whether we should appeal to the color-octet scheme, or to what extent it manifests itself in the description of quarkonium production is still a challenging question.
A number of investigations indicate that the leading-order(LO) QCD calculation are inadequate to explain the experimental data. It turns out that some of the discrepancies between LO QCD calculation and experimental observation can be remedied by including higher order corrections, e.g. the double charmonium production at $B$ factories \cite{bralee,hqiao,ZYJ,zyjchao,gbwang,GB1,qiao}, but in some other cases the color-octet contributions are still indispensable, where the QCD higher order calculation is normally hard to proceed. To consider higher order contributions, including perturbative QCD and relativistic corrections,
tends to be an inevitable task in the investigation of heavy quarkonium production.

Quarkonium production in $\gamma\gamma$ collision is an interesting topic to study, where the signals are relatively clean. In 2001, the DELPHI Collaboration reported its measurement on $J/\psi$ inclusive production via photon-photon interaction in the run of LEP\uppercase\expandafter{\romannumeral2} \cite{delphi}.
Based on leading order NRQCD analyses, Klasen {\it et al.} \cite{Kniehl} found that the DELPHI data evidently favor the NRQCD color-octet(CO) mechanism, but rather merely the conventional color-singlet(CS) model. However, it should be noted that in \cite{Kniehl} the superficially higher order process $\gamma+\gamma\rightarrow J/\psi+c+\bar{c}$ was not considered, but in fact later found is the dominant CS process \cite{qiaow}.
Despite this, the great discrepancy between CS prediction and experiment data still remains. Nevertheless, before filling the gap by color-octet contributions, we should stretch our CS calculation as far as we can.
To this aim, we calculate the tedious next-to-Leading order(NLO) QCD corrections to $\gamma+\gamma\rightarrow J/\psi+c+\bar{c}$ process in this work, the first true NLO calculation of 2 to 3 inclusive process in heavy quarkonium production.

\begin{figure}[thbp]
\begin{center}
\includegraphics[scale=0.7]{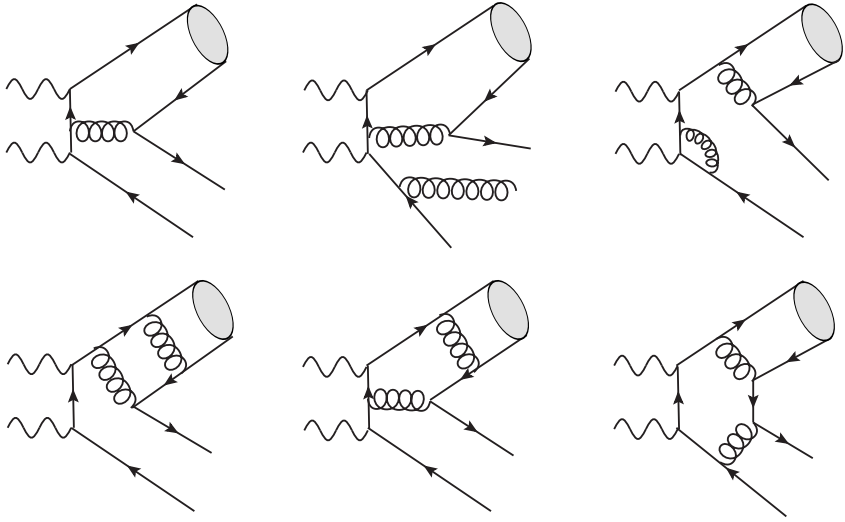}
\caption{Typical Feynman diagrams for $\gamma+\gamma\rightarrow
J/\psi+c+\bar{c}+X$ process at LO and NLO . \label{fig1}}
\end{center}
\end{figure}

In the calculation, the initial photons are assumed to be generated by the bremsstrahlung in high-energy electron-positron collision. The source of photons can be well formulated in the Weizsacker-Williams(WW) approximation as \cite{WWA}:
\bqa
f_\gamma(x)=\frac{\alpha}{2\pi}(\frac{1+(1-x)^2}{x}log(\frac{Q_{max}^2}{Q_{min}^2})+
2m_e^2x(\frac{1}{Q_{max}^2}-\frac{1}{Q_{min}^2}))\ ,
\label{dis}
\eqa
where $Q_{min}^2=m_e^2x^2/(1-x)$ and $Q_{max}^2=(\theta_c\sqrt{s}/2)^2(1-x)+Q_{min}^2$ with $x=E_\gamma/E_e$, $\theta_c$ the maximum scattering angle of electron and positron.

The total cross section of $J/\psi$ production in photon-photon collision can be obtained by convoluting the $\gamma+\gamma\rightarrow J/\psi+c+\bar{c}$ subprocess with the photon distribution functions, i.e.,
\bqa
d\sigma=\int{dx_1dx_2f_\gamma(x_1)f_\gamma(x_2)d \hat{\sigma}(\gamma + \gamma\rightarrow J/\psi + c + \bar{c})}\ .
\eqa
Here, the CS subprocess is calculated in NRQCD framework to the NLO, which can be schematically formulated as
\bqa
d\hat{\sigma}(\gamma+\gamma\rightarrow J/\psi+c+\bar{c})=d\hat{\sigma}_{born}+d\hat{\sigma}_{virtual}+ d\hat{\sigma}_{real}+\mathcal{O}(\alpha^2\alpha_s^4)\ .
\label{subcs}
\eqa
In (\ref{subcs}) the born level cross section, virtual and real corrections take the following forms:
\bqa
\begin{split}
&d\hat{\sigma}_{born}=\frac{1}{2\Phi} \overline{\sum}|\mathcal{M}_{born}|^2dPS_3\ ,\\
&d\hat{\sigma}_{virtual}=\frac{1}{2\Phi} \overline{\sum}2Re(\mathcal{M}^*_{born}\mathcal{M}_{NLO})dPS_3\ ,\\
&d\hat{\sigma}_{real}=\frac{1}{2\Phi} \overline{\sum}|\mathcal{M}_{real}|^2dPS_4\ .
\end{split}
\eqa
Here $\Phi=\lambda^{1/2}(s,0,0)$ is the flux; $\overline{\sum}$ means sum (average) over the polarizations of final (initial) state particles; $dPS_3$ and $dPS_4$ denote for three- and four-body phase spaces, respectively.

The standard form of quarkonium spin projection operator is adopted in our calculation \cite{pchoaleb,bodwpe,bralee}. For color-singlet and spin-triplet quarkonium, it reads
\bqa
v(\bar{p})\bar{u}(p) =\frac{1}{4\sqrt{2} E(E+M)}(\not\!\bar{p}-m_c)
\not\!\epsilon_S^*(\not\!P+2E) (\not\!p+m_c)
\otimes\bigg(\frac{\bf{1}_c} {\sqrt{N_c}}\bigg)\ .
\label{projector}\eqa
Here, $p$ and $\bar{p}$
are the momenta of quark and antiquark respectively;
$\epsilon^\mu_S$ is the polarization vector of the quarkonium, $P=p+\bar{p}$ denotes its four-momentum, and $M=2m_c$ is the mass of quarkonium; $E^2=P^2/4$, $N_c=3$, and $\bf{1}_c$ represents the unit
color matrix. For color-singlet and spin-singlet state, the
projection operator could be obtained by replacing the $\not\!\epsilon_S^*$
in Eq.(\ref{projector}) by a $\gamma_5$. In the case of $J/\psi$ production, for leading order NRQCD relative velocity expansion, we are legitimate to take $p = \bar{p} = P/2$.

The calculation is performed in the Feynman gauge, and the conventional dimensional regularization with $D = 4-2\epsilon$ is adopted in regularizing the ultraviolet and infrared divergences.
In the end of the day, all ultraviolet divergences are completely canceled by renormalization, while the Coulombic singularities are factorized out and attributed to the NRQCD long-distance matrix elements. Infrared divergences arising from loop integration are canceled by including the real emission processes and renormalization terms.

In the calculation of the real corrections, the phase space slicing method \cite{twocut} is adopted in order to separate the infrared singularities. The outgoing gluon with energy $p_g^0<\delta$ is considered to be soft, while $p_g^0>\delta$ is taken as a hard one. Then, the real corrections are factorized as
\bqa
d\sigma_{real}=d\sigma_{soft}^{IR}|_{p_g^0<\delta}+ d\sigma_{hard}^{IR-free}|_{p_g^0>\delta}\ ,
\eqa
where the soft divergent term and finite part are properly separated.

The ultraviolet and infrared divergences also exist in the
renormalization constants $Z_2, Z_3, Z_m$, and $Z_g$, which correspond
respectively to the quark field, gluon field, quark mass, and strong
coupling constant $\alpha_s$. Among them, $Z_g$ is defined in the modified-minimal-subtraction
$(\overline{MS})$ scheme, while the other three are in the on-shell (OS)
scheme. Therefore, we can readily get the corresponding counter terms,
\begin{eqnarray}
\delta Z_2^{\rm OS}&=&-C_F\frac{\alpha_s}{4\pi}
\left[\frac{1}{\epsilon_{\rm UV}}+\frac{2}{\epsilon_{\rm IR}}
-3\gamma_E+3\ln\frac{4\pi\mu^2}{m^2}+4\right],
\nonumber\\
 \delta Z_m^{\rm OS}&=&-3C_F\frac{\alpha_s}{4\pi}
\left[\frac{1}{\epsilon_{\rm
UV}}-\gamma_E+\ln\frac{4\pi\mu^2}{m^2} + \frac{4}{3}\right], \nonumber\\
 \delta
Z_3^{\mathrm{OS}}&=&\dfrac{\alpha_s}{4\pi} \biggl[(\beta_0-2C_A)(\dfrac{1}{\epsilon_{UV}}
-\dfrac{1}{\epsilon_{IR}}) -\dfrac{4}{3}T_F(\dfrac{1}{\epsilon_{UV}} -\gamma_E +\ln\dfrac{4\pi \mu^2}{m^2}) \biggr],
 \nonumber\\
  \delta Z_g^{\overline{\rm MS}}&=&-\frac{\beta_0}{2}\,
  \frac{\alpha_s}{4\pi}
  \left[\frac{1}{\epsilon_{\rm UV}} -\gamma_E + \ln(4\pi)
  \right]\ .
\end{eqnarray}

In the calculation, \textbf{FeynArts} \cite{feynarts} was used to generate the Feynman diagrams, as shown in FIG.1. Among them, there are 20 tree-level diagrams, 442 one-loop virtual correction diagrams, and 200 real radiation ones. Apart from these, we also need 140 counter diagrams for the renormalization. Self-written codes based on \textbf{FeynCalc} \cite{Feyncalc} are developed to apply the spin projector and to perform the tensor reduction. The package \textbf{Fire} \cite{fire} is employed to reduce all one-loop integrals to typical
master-integrals $A_0$, $B_0$, $C_0$, and $D_0$, and which are numerically evaluated by the package \textbf{LoopTools} \cite{looptools}.
In the end, the overall integrals are performed numerically with the assistance of package \textbf{CUBA} \cite{cuba}. Note, in the course of calculation, the packages \textbf{\$Apart}\cite{apart} and \textbf{FeynCalcFormLink}\cite{formlink} are sometimes used to facilitate the process.

In the numerical calculation, the inputs are commonly used and from the experiment measurement, i.e., $\alpha=1/137.065$, $m_c=1.5\pm0.1$ GeV and $m_e=0.511$ MeV. The radial wave function at the origin of $J/\psi$ is
extracted from the $J/\psi\rightarrow e^+e^-$ process with the latest PDG data $\Gamma(J/\psi\rightarrow e^+e^-)=5.55\pm 0.14\pm 0.02$ keV \cite{PDG}, i.e. $|R_s(0)|^2=1.01$ GeV$^3$. The renormalization and the factorization scales are set to be the same, i.e. $\mu= \mu_r=\mu_f=r\sqrt{4m_c^2+p_t^2}$. Here $p_t$ is the transverse momentum of $J/\psi$, and $r=\{0.5,1,2\}$. The two-loop formula for the running coupling constant $\alpha_s(\mu)$ is employed to gauge its scale dependence,
\beq \frac{\alpha_s(\mu)}{4\pi}=\frac{1}{\beta_0 L}-\frac{\beta_1\ln
L}{\beta_0^3L^2}\ . \eeq
Here, $L=\ln(\mu^2/\Lambda_{QCD}^2)$,
$\beta_0=(11/3)C_A-(4/3)T_Fn_f$, and
$\beta_1=(34/3)C_A^2-4C_FT_Fn_f-(20/3)C_AT_Fn_f$, with
non-perturbative QCD cutoff $\Lambda_{QCD}$ and active quark number $n_f$ set to be \cite{PDG}
\beq
\begin{split}
&1.5\ \text{GeV}<\mu\le4.7\ \text{GeV},\ \ n_f=4,\ \ \Lambda_{QCD}=297\ \text{MeV},\\
&4.7\ \text{GeV} < \mu\le173\ \text{GeV},\ \ n_f=5,\ \ \Lambda_{QCD}=214\ \text{MeV}.\\
\end{split}
\eeq

Some other parameters pertaining to LEP\uppercase\expandafter{\romannumeral2} experiment take as follows: the collision energy $\sqrt{s}=197$ GeV; the angle cut $\theta_c=32$ mrad in Eq.(\ref{dis}); the rapidity lies in $-2<y<2$; and the invariant mass of the two photon system is constrained as $W\le35$ GeV in order to exclude non-photoproduction processes in our calculation. Note that the feeddown factor $0.278$ of $\psi^{'}\rightarrow J/\psi+X$ to the concerned process is also included in our numerical calculation, which is deduced from the branching fraction and production rate of $\psi^{'}$.

In Fig.2, the $J/\psi$ transverse momentum squared distributions through $\gamma+\gamma\rightarrow J/\psi+c+\bar{c}$ process are illustrated for LEP\uppercase\expandafter{\romannumeral2} experimental environment.
The DELPHI measurement \cite{delphi} and contributions of other subprocesses \cite{Kniehl} are shown as well for comparison. We notice from the plot that with the increase of $p_t$, the NLO correction decreases and turns to be negative at about 8 GeV$^2$. And, the tremendous discrepancy between experimental data and CS model prediction still remains even with the NLO QCD correction for $\gamma+\gamma\rightarrow J/\psi+c+\bar{c}$ process, the dominant one.
In this sense, the CO mechanism is still necessary to explain the experimental data.

\begin{figure}
\centering
\includegraphics[scale=0.7]{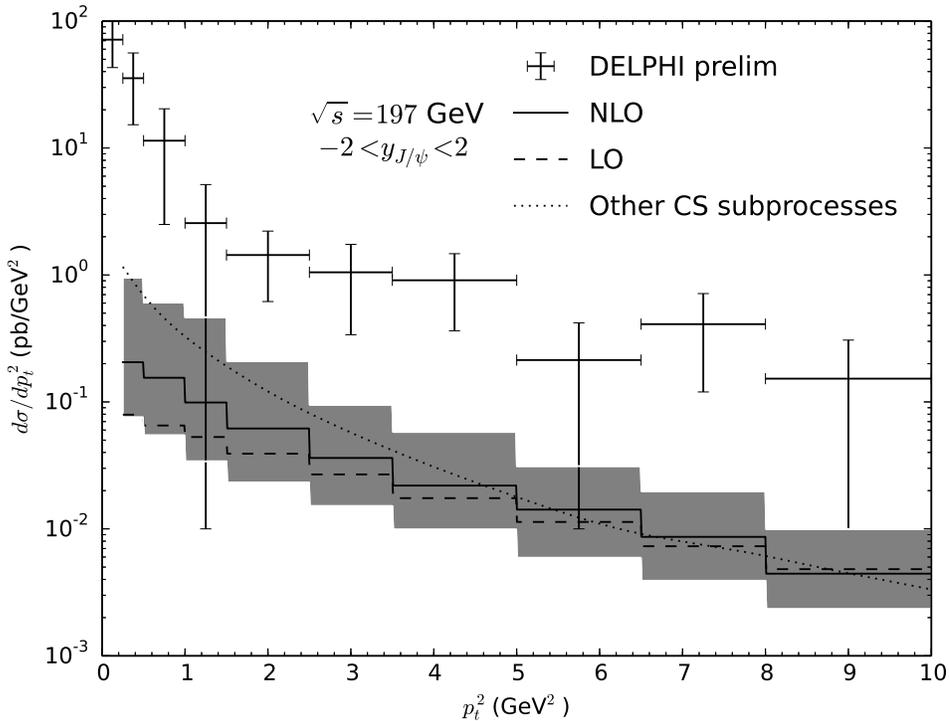}
\caption{The $p_t^2$ distributions of $J/\psi$ production in $\gamma\gamma$ collision at LEP\uppercase\expandafter{\romannumeral2} environment. The shaded band represents the NLO result with upper bound for $r=0.5$, $m_c=1.4$ GeV and lower bound for $r=2$, $m_c=1.6$ GeV. The solid and dashed lines denote the NLO and LO results respectively with $r=1$ and $m_c=1.5$ GeV.}
\end{figure}

The LO and NLO total cross sections within the integrated transverse momentum range $1\le p_t^2\le10$ GeV$^2$ are given in Table.\uppercase\expandafter{\romannumeral1}.
In order to show the influence of the renormalization scale and charm quark mass, we chose $m_c=1.4,1.5,1.6$ GeV and $r=0.5,1,2$ respectively in the numerical evaluation. And find that the $K$ factor at the central values of these parameters $r$ and $m_c$ is about 1.46, the uncertainties induced by $m_c$ and $r$ are even larger at NLO than that at LO. This is due to the fact that in small interval $p_t^2$---$p_t^2+\Delta p_t^2$ the NLO cross section can be expressed in form of
\begin{equation}
\hat{\sigma}_{NLO}|_{\Delta p_t^2}= \hat{\sigma}_{LO}|_{\Delta p_t^2}(1 + (\alpha_s\xi)|_{\Delta p_t^2})\ .
\end{equation}
The factor $\xi$ and the strong coupling constant $\alpha_s$ increase with the decrease of $r$ and $m_c$ mostly in the region of $1\le p_t^2\le10$ GeV$^2$, which eventually leads to a relatively large uncertainties at NLO.
We notice that the DELPHI measurement gives $(6.4\pm2.0)$ pb for $J/\psi$ inclusive production in electron-position collision, while CS processes other than our concern only yields $(0.39_{-0.09}^{+0.16})$ pb \cite{Kniehl}.
The total CS prediction after including the dominant process of
$\gamma+\gamma\rightarrow J/\psi+c+\bar{c}$ with NLO QCD correction is still 3 times smaller than the experimental data. Therefore, we may reasonably infer that to explain the $J/\psi$ inclusive production data at LEP, merely color-singlet mechanism is not enough.

\begin{table}[h]
\caption{The NLO(LO) results of total cross sections with different renormalization scales and charm quark masses. The LO result agrees with what in Refs. \cite{LiRong,SunZhan} after taking the same inputs.}
\begin{center}
\renewcommand\arraystretch{1}
  \begin{tabular}{p{1.5cm}*{3}{c}}
    \toprule
    $\ \ \sigma$(pb)& $m_c=1.4$ GeV & $m_c=1.5$ GeV & $m_c=1.6$ GeV\\
    \hline
    $r=0.5$ &
    $\quad 0.766(0.436)\quad$ &
    $\quad 0.459(0.283)\quad$ &
    $\quad 0.299(0.187)\quad$ \\
    $r=1$ &
    $\quad 0.363(0.236)\quad$ &
    $\quad 0.227(0.156)\quad$ &
    $\quad 0.152(0.105)\quad$ \\
    $r=2$ &
    $\quad 0.216(0.152)\quad$ &
    $\quad 0.138(0.101)\quad$ &
    $\quad 0.093(0.069)\quad$ \\
    \botrule
  \end{tabular}
\end{center}
\label{tab:cro1}
\end{table}

In the future electron-positron colliders, like CEPC run at $\sqrt{s}=250$ GeV, the $J/\psi$ production through $\gamma\gamma$ collision can be measured with high accuracy. Schematically, we also extend our research on $\gamma+\gamma\rightarrow J/\psi+c+\bar{c}$ process to colliding energy of $\sqrt{s}=250$ GeV. The Weizsacker-Williams approximation is applied to the initial photons with the same $\theta_c$ as in the discussion of LEP\uppercase\expandafter{\romannumeral2} case, while the constraint to the center-of-mass energy of two initial photons is not employed.
Fig.3 shows the NLO and LO $p_t^2$ distributions of $J/\psi$ production at $\sqrt{s}=250$ GeV. The differential cross section is significantly enhanced by NLO corrections, especially in low-$p_t$ region.
Integrating over the $p_t^2>1$ GeV$^2$ region, we then get the total NLO(LO) cross section as 0.432(0.245) pb. The $K$ factor here is about 1.76, quite larger than that at LEP\uppercase\expandafter{\romannumeral2} energy.

\begin{figure}
\centering
\includegraphics[scale=0.7]{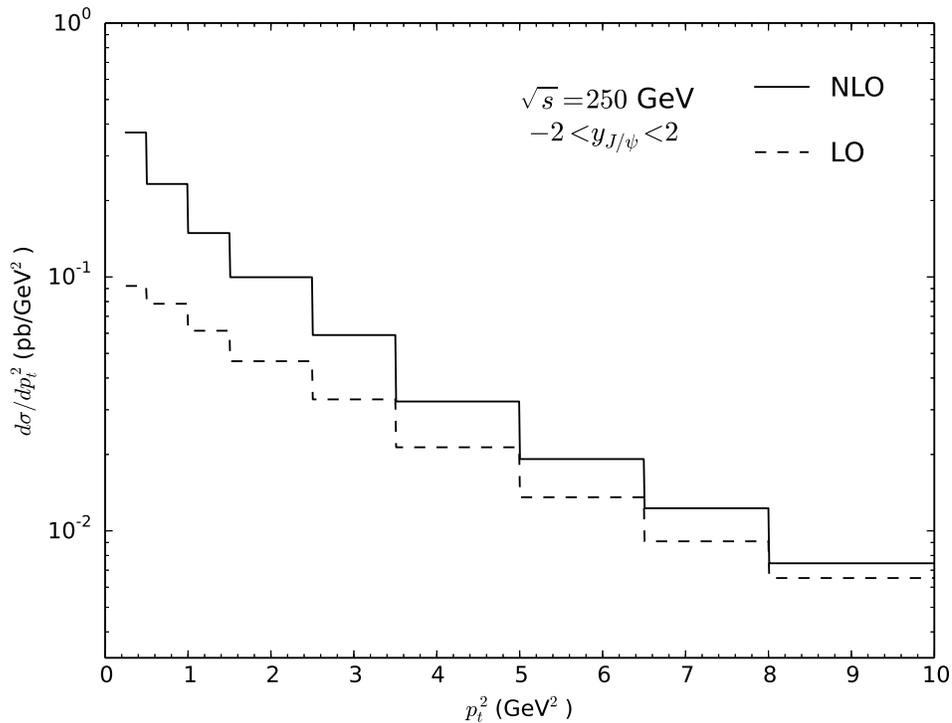}
\caption{The $p_t^2$ distribution of $J/\psi$ production through $\gamma\gamma$ collision at CEPC. The solid and dashed lines represent the NLO and LO results with $r=1$ and $m_c=1.5$ GeV.}
\end{figure}

In conclusion, we have calculated the NLO QCD corrections to $\gamma+\gamma\rightarrow J/\psi+c+\bar{c}$ process. The numerical evaluation aims at the LEP\uppercase\expandafter{\romannumeral2} experiment for illustration, in which the transverse momentum distribution is displayed, the total cross section is calculated, and the theoretical uncertainties are estimated. Result shows that for LEPII experiment, the NLO QCD corrections yield a $K$ factor of 1.46, which indicates that the huge discrepancy between CS prediction and experimental observation remains and the CO contributions are still indispensable. The numerical evaluation is also performed at the future electron-positron collider, the CEPC, energy, and we find the NLO total cross section(0.432 pb) is about 1.76 times larger than the LO result(0.245 pb).

It is notable that in our calculation the initial photons are thought to be generated by the bremsstrahlung in electron-positron collision, which is well described by WW approximation. In the condition of high collision energy, more energetic and luminous photons can also be generated by laser back scattering(LBS) mechanism \cite{LBS}, which will make a real photon collider.

%%%%%%%%%%%%%%%%%%%%%%%%%%%%%%%%%%%%%%%%%%%%%%%%%%%%%%%%%%%%%%%%%%%%%
\vspace{0.6cm} {\bf Acknowledgments}

We appreciate Rong Li's helpful discussion on the leading order result. This work was supported in part by the Ministry of Science and Technology of the People¡¯s Republic of China (2015CB856703), by by the Strategic Priority Research Program of the Chinese Academy of Sciences (XDB23030000), and by the National Natural Science Foundation of China(NSFC) under the grant 11375200.
%%%%%%%%%%%%%%%%%%%%%%%%%%%%%%%%%%%%%%%%%%%%%%%%%%%%%%%%%%%%%%%%%%%%


\vspace{0.7cm}
\begin{thebibliography}{99}
%---------------------------------------------------------------
\bibitem{NRQCD}  G.~T.~Bodwin, E.~Braaten and G.~P.~Lepage,
  %``Rigorous QCD analysis of inclusive annihilation and production of heavy quarkonium,''
  Phys.\ Rev.\ D {\bf 51}, 1125 (1995)
  [Erratum-ibid.\ D {\bf 55}, 5853 (1997)].
%---------------------------------------------------------------
\bibitem{bralee}  E.~Braaten and J.~Lee,
  %``Exclusive double charmonium production from e+ e- annihilation into a virtual photon,''
  Phys.\ Rev.\ D {\bf 67}, 054007 (2003)
  [Erratum-ibid.\ D {\bf 72}, 099901 (2005)].
%---------------------------------------------------------------
\bibitem{hqiao} K.~Hagiwara, E.~Kou and C.~F.~Qiao,
  %``Exclusive $J/\psi$ productions at $e^{+} e^{-}$ colliders,''
  Phys.\ Lett.\ B {\bf 570}, 39 (2003).
%---------------------------------------------------------------
\bibitem{ZYJ} Y.~J.~Zhang, Y.~j.~Gao and K.~T.~Chao,
  %``Next-to-leading order QCD correction to e+ e- ---> J / psi + eta(c) at s**(1/2) = 10.6-GeV,''
  Phys.\ Rev.\ Lett.\  {\bf 96}, 092001 (2006).
%---------------------------------------------------------------
\bibitem{zyjchao} Y.~J.~Zhang and K.~T.~Chao,
  %``Double charm production e+ e- ---> J / psi + c anti-c at B factories with next-to-leading order QCD correction,''
  Phys.\ Rev.\ Lett.\  {\bf 98}, 092003 (2007).
%---------------------------------------------------------------
\bibitem{gbwang} B.~Gong and J.~X.~Wang,
  %``Next-to-leading-order QCD corrections to e+e- --> J/psi(cc) at the B factories,''
  Phys.\ Rev.\ D {\bf 80}, 054015 (2009).
\bibitem{GB1} B.~Gong and J.~X.~Wang,
  %``QCD corrections to $J/\psi$ plus $\eta_c$ production in $e^{+} e^{-}$ annihilation at $S^{(1/2)}$ = 10.6-GeV,''
  Phys.\ Rev.\ D {\bf 77}, 054028 (2008).
%---------------------------------------------------------------
\bibitem{qiao} L.~B.~Chen and C.~F.~Qiao,
  %``NLO corrections to $\chi_{bJ}$ to two-$J/\psi$ exclusive decay processes,''
  Phys.\ Rev.\ D {\bf 89}, 074004 (2014).

%---------------------------------------------------------------
\bibitem{delphi} S.Todorova-Nova, in Proceedings of the XXXI International Symposium on Multiparticle Dynamics, Datong, China, 2001(World Scientific, Singapore, to be published), hep-ph/0112050; M.Chapkin, in Proceedings of the 7th International Workshop on Meson Production, Properties and Interaction(Meson 2002), Krakow, Poland, 2002(World Scientific, Singapore, to be published).

\bibitem{Kniehl} Michael Klasen {\it et al.}, Phys. \ Rev.\ Lett.\ {\bf 89}, 032001(2002).

%------------------------------------------------------------
\bibitem{qiaow} C.~F.~Qiao and J.~X.~Wang, Phys.\ Rev.\ D {\bf 69}, 014015(2004).

\bibitem{WWA} S.Frixione, M.L.Mangano, P.Nason, and G.Ridolphi, Phys.Lett.B {\bf319},339(1993).
%---------------------------------------------------------------
\bibitem{pchoaleb} P.~L.~Cho and A.~K.~Leibovich,
  %``Color octet quarkonia production,''
  Phys.\ Rev.\ D {\bf 53}, 150 (1996).
%----------------------------------------------------------------
\bibitem{bodwpe} G.~T.~Bodwin and A.~Petrelli,
  %``Order $v^4$ corrections to $S$ wave quarkonium decay,''
  Phys.\ Rev.\ D {\bf 66}, 094011 (2002)
  [Erratum-ibid.\ D {\bf 87}, no. 3, 039902 (2013)].
%----------------------------------------------------------------
\bibitem{twocut} B.~W.~Harris and J.~F.~Owens,
  %``The Two cutoff phase space slicing method,''
  Phys.\ Rev.\ D {\bf 65}, 094032 (2002).
%----------------------------------------------------------------
\bibitem{feynarts} T.~Hahn,
  %``Generating Feynman diagrams and amplitudes with FeynArts 3,''
  Comput.\ Phys.\ Commun.\  {\bf 140}, 418 (2001).
%----------------------------------------------------------------
\bibitem{Feyncalc} R.~Mertig, M.~Bohm and A.~Denner,
  %``FEYN CALC: Computer algebraic calculation of Feynman amplitudes,''
  Comput.\ Phys.\ Commun.\  {\bf 64}, 345 (1991).
  %----------------------------------------------------------------
\bibitem{fire} A.~V.~Smirnov,
  %``Algorithm FIRE -- Feynman Integral REduction,''
  JHEP {\bf 0810}, 107 (2008).
%----------------------------------------------------------------
\bibitem{looptools} T.~Hahn and M.~Perez-Victoria,
  %``Automatized one loop calculations in four-dimensions and D-dimensions,''
  Comput.\ Phys.\ Commun.\  {\bf 118}, 153 (1999).
  %----------------------------------------------------------------
\bibitem{cuba} T.~Hahn,
  %``Generating Feynman diagrams and amplitudes with FeynArts 3,''
  Comput.\ Phys.\ Commun.\  {\bf 168}, 78 (2005).
  %----------------------------------------------------------------
\bibitem{apart}F.~Feng,
  %``$Apart: A Generalized Mathematica Apart Function,''
  Comput.\ Phys.\ Commun.\  {\bf 183}, 2158 (2012).
\bibitem{formlink} F. Feng, arXiv:1212.3522.
%----------------------------------------------------------------
\bibitem{PDG} J.~Beringer {\it et al.}  [Particle Data Group Collaboration],
  %``Review of Particle Physics (RPP),''
  Phys.\ Rev.\ D {\bf 86}, 010001 (2012).
  %K.~A.~Olive {\it et al.}  [Particle Data Group Collaboration],
  %``Review of Particle Physics,''
  %Chin.\ Phys.\ C {\bf 38}, 090001 (2014).
%----------------------------------------------------------------
\bibitem{LiRong} R. Li and K. T. Chao, Phys.\ Rev.\ D {\bf 79}, 114020 (2009).
\bibitem{SunZhan} Z. Sun, X. G. Wu and H. F. Zhang, Phys.\ Rev.\ D {\bf 92}, 074021 (2015).
\bibitem{LBS} I.F.Ginzburg, G.L.Kotkin, V.G.Serbo and C.I.Telnov, Nucl. Instrum. Methods Phys. Res. {\bf205}, 47 (1983)
\end{thebibliography}
\end{document}